\documentclass[letterpaper,journal]{IEEEtran}

\usepackage{amsmath,amssymb,amsfonts}
\usepackage{array}
\usepackage{booktabs}
\usepackage{tabularx}
\usepackage{cite}
\usepackage{graphicx}
\usepackage{microtype}
\usepackage{url}

\newcommand{\system}{\textsc{ExplainRoute}}
\newcommand{\auditcase}[1]{\texttt{#1}}

\begin{document}

\title{ExplainRoute: A Pre-Deployment Audit Framework for Non-Answer-Giving Programming Tutors}

\author{Yiming Gai, Yingying Zhang and Xuefei Huang}

\maketitle

\begin{abstract}
Programming tutors should make room for a learner's own explanation rather than immediately replacing it with a model answer. We present \system as a pre-deployment audit framework for a non-answer-giving teaching assistant: it reads a code line and a learner explanation, estimates the explanation state, and selects one of two bounded responses---a Feynman-style open self-explanation prompt or a Socratic scaffold. The framework exposes its state, strategy, cited code fragment, and leakage risk in a machine-checkable contract. Unlike a benchmark that ranks tutors by a single fluency score, it tests information boundaries, response polarity, failure closure, and the value of learner-explanation visibility before a classroom study. We evaluate it offline on the 1,770-pair SelfCode corpus. A code-group split leaves 443 pairs in 11 untouched holdout groups until generation is sealed. The holdout compares a direct answer, fixed open self-explanation, fixed Socratic scaffolding, adaptive two-role routing, and an adaptive no-state ablation. Contract validity is 100\% for both fixed pedagogical conditions and the adaptive condition; adaptive routing agrees with the frozen score-based reference rule on 60.5\% of records. Its state macro-F1 is 0.238, falling between the two fixed strategies (Open 0.229; Socratic 0.246) and showing no reliable adaptive advantage. An independent language-model judge rates adaptive responses 4.516/5 on a composite rubric, above the no-state ablation (2.819/5) but slightly below fixed open self-explanation (4.598/5) and Socratic scaffolding (4.658/5). A blinded teaching-assistant rubric evaluation on a stratified 40-row subset corroborates the information-value result (a visible learner explanation beats the no-state ablation) and the absence of an adaptive advantage over fixed strategies. The resulting contribution is a validated audit protocol and a boundary finding, not a claim of improved learning, retention, or causal instructional effectiveness.
\end{abstract}

\begin{IEEEkeywords}
teaching-assistant agents, programming education, self-explanation, Socratic scaffolding, Feynman technique, learner-state estimation, auditable LLM systems
\end{IEEEkeywords}

\section{Introduction}
\IEEEPARstart{P}{rogramming} learners can recognize that a line of code ``does something'' without being able to reconstruct why it does so. A teaching assistant that immediately supplies the correct interpretation may produce a fluent exchange while hiding the learner's mental model and reducing the opportunity to repair it. Self-explanation asks the learner to reconstruct an explanation in the learner's own words, making missing links and misconceptions observable \cite{chi1994selfexplanation,graesser1994construction,chapagain2023selfcode}. Socratic tutoring uses a question to elicit the missing relation without disclosing the conclusion \cite{graesser2001autotutor,vanlehn2011relative}. These strategies are complementary, but a deployed tutor needs an observable rule for deciding which response to produce.

Large language models have made conversational tutoring easy to prototype \cite{kasneci2023chatgpt,denny2023computing,prather2023robots}, while making evidence acquisition and response decisions difficult to audit. Agentic systems address this opacity by exposing plans, tools, intermediate decisions, and verification actions \cite{yao2023react,shinn2023reflexion,madaan2023selfrefine}. However, adding roles does not guarantee a better tutor: a longer workflow can introduce cost, routing errors, or answer leakage. A rigorous teaching-assistant evaluation should therefore compare adaptive routing against fixed strategies, retain failure traces, and distinguish response quality proxies from learning outcomes \cite{holmes2019ai,luckin2016intelligence}.

This paper studies that bounded pre-deployment audit problem. \system\ is a two-role teaching-assistant agent: a code-focused diagnostician produces a structured explanation-state record, and a tutor role chooses a single Feynman-style or Socratic action. A deterministic contract rejects malformed or answer-leaking scaffold outputs. The experiment uses SelfCode, an annotated corpus of learner and expert explanations of Java lines \cite{chapagain2023selfcode}. Expert fields remain hidden until generation artifacts are sealed. The paper is not another attempt to obtain the highest aggregate score from an LLM tutor. Its unit of analysis is the auditable interaction: what information was visible, which action was selected, whether the response obeyed the pedagogical contract, and where adaptive routing failed. We ask:
\begin{enumerate}
\item \textbf{RQ1:} Can a non-answer-giving programming tutor be constrained by a machine-checkable contract that exposes its state estimate, action, code evidence, and leakage risk?
\item \textbf{RQ2:} Does observing the learner's explanation improve observable tutoring-response quality relative to a no-state ablation?
\item \textbf{RQ3:} Does adaptive routing outperform fixed open or Socratic strategies on offline proxy endpoints, or does the audit reveal that the adaptive state estimate is not yet strong enough? We pre-register both outcomes as informative.
\end{enumerate}

Our contributions are: (i) a pre-deployment audit framework for non-answer-giving programming tutors, with explicit state, strategy, code evidence, leakage fields, and fail-closed response polarity; (ii) a frozen group-held-out evaluation whose five conditions isolate answer-giving, fixed pedagogy, learner-explanation visibility, and adaptive routing; (iii) a two-evaluator analysis that triangulates an independent rubric judge with blinded teaching assistants; and (iv) a reproducible artifact bundle with generation seals, failure logs, aggregate results, and pre-registered follow-up protocols. The paper deliberately makes no claim about student learning gain. Its negative RQ3 result is treated as a design finding: an audit can show that an apparently adaptive route should not yet be over-deployed.

\section{Related Work}
\subsection{Self-explanation and programming education}
Self-explanation is one of the most robust learning strategies in the cognitive-science literature: learners who generate their own explanations of worked material produce more inferences, connect new information to prior knowledge, and repair faulty mental models more effectively than learners who merely study the material \cite{chi1994selfexplanation,chi2000selfexplanation}. The benefit generalizes across domains and age groups, and self-explanation can be taught as a metacognitive strategy that learners then transfer to new material \cite{bielaczyc1995training,aleven2002effective,roy2005self}. Learning from errors and from one's own incorrect inferences is itself a distinct mechanism that complements correct-case self-explanation \cite{metcalfe2017learning}. A consistent finding across this work is that the act of articulating an explanation---not merely reading a correct one---is what drives understanding, which motivates tutoring designs that elicit, rather than replace, the learner's explanation.

Programming gives this process a specific structure. Program-comprehension research shows that even novice programmers construct mental models of code at both a surface, syntactic level and a deeper plan-based level, and that misconceptions often arise from gaps between the two \cite{pennington1987stimulus,soloway1984learning,ramalingam2004development}. SelfCode operationalizes this gap as a corpus: it pairs 1,770 crowd-sourced learner explanations of Java code lines with expert explanations and five-point semantic-similarity judgments, providing an observable proxy for explanation quality \cite{chapagain2023selfcode}. We use SelfCode as an offline benchmark for recognizing explanation state, not as evidence of a treatment effect. Relative to prior self-explanation prompting in programming and other domains, our contribution is a bounded, auditable routing protocol whose action is constrained to remain non-answer-giving.

\subsection{Intelligent tutors and learner modeling}
Classical intelligent tutoring systems (ITS) decompose instruction into three components---a domain model, a learner (student) model, and a pedagogical strategy---and a substantial body of work shows that such systems can approximate one-on-one human tutoring, which remains the effectiveness ceiling for instruction \cite{anderson1985intelligent,graesser2001autotutor,vanlehn2011relative}. AutoTutor in particular demonstrated that Socratic, expectation-driven dialogue can elicit deeper reasoning from learners without simply handing them the answer \cite{graesser2001autotutor}. The explanatory gap between what a learner states and what an expert expects is precisely what such a tutor must diagnose and then act on.

Modern learner models make those inferences probabilistic: Bayesian student models and knowledge-tracing methods maintain uncertainty over latent knowledge components and use it to select the next instructional step \cite{conati2002probabilistic,du2019knowledge}. ExplainRoute inherits the same diagnose-then-act structure, but deliberately restricts the state space to four values because SelfCode supplies semantic-similarity labels rather than a complete skill model. We therefore report the state mapping and its empirical limitations instead of presenting it as instructor ground truth, and we keep the human-designable surface---dataset, prompt, schema, and budgets---explicit so that the agent's decisions remain auditable.

\subsection{LLM agents, retrieval, and tutoring evaluation}
The recent wave of LLM agent frameworks makes intermediate reasoning and tool use explicit: ReAct interleaves reasoning traces with actions, Toolformer and Gorilla equip models with external tools, and Self-Refine and Reflexion add iterative feedback loops that improve generated artifacts \cite{yao2023react,schick2023toolformer,patil2023gorilla,madaan2023selfrefine,shinn2023reflexion}. Chain-of-thought and plan-and-solve prompting further show that eliciting explicit reasoning steps can change model behavior on complex tasks \cite{wei2022chain,kojima2022large,wang2023plan}. These properties are attractive for tutoring because they make a response chain inspectable, but they also introduce cost, routing errors, and answer-leakage risk that a teaching assistant must bound. ExplainRoute keeps the agent to two roles with a deterministic verifier and no post-hoc repair, so that every decision is logged and every failure is retained.

A parallel literature examines generative AI in computing education, where the central tension is that a model fluent enough to explain code can also short-circuit the learning process by supplying answers \cite{kasneci2023chatgpt,denny2023computing,prather2023robots,denny2024genai}. Retrieval-augmented generation and self-reflective retrieval improve grounding in open-domain settings \cite{lewis2020rag,gao2023rag,asai2024selfrag,mainrag2025,ragcritic2025}, but retrieval quality is not the focus of our experiment; we intentionally omit retrieval so that any difference between conditions can be attributed to learner-explanation visibility and response routing rather than to evidence acquisition. Finally, recent LLM-tutor benchmarks separate grounding, pedagogical action, and safety instead of collapsing them into a single fluency score \cite{lee2025mathtutorbench,wang2025mrbench}; our independent rubric follows this separation and treats its composite score as an observable quality proxy rather than a learning outcome.

\subsection{Auditable tutoring evaluation}
ExplainRoute is related to LLM-tutor benchmarks, but it addresses a different pre-deployment question. A benchmark typically ranks systems by response quality on a fixed prompt set; an audit framework first asks whether the system had access to the permitted information, selected an allowed pedagogical action, cited the supplied code, and failed closed when the contract was violated. This distinction matters for non-answer-giving tutoring: a fluent response can still disclose the answer, misread the learner, or conceal an unverified routing decision. Our shared schema, deterministic verifier, generation seal, and retained failure traces are therefore part of the contribution, not implementation details added after scoring.

The five-condition design operationalizes this distinction. Direct isolates the actionability and leakage tradeoff of giving the answer. Fixed Open and Fixed Socratic measure the two bounded pedagogical actions without adaptive routing. No-State removes the learner explanation while retaining the code line, isolating the information value of observing the learner. Adaptive adds the diagnostic role and tests whether state-aware routing improves on those fixed actions. All five conditions use the same response schema and main tutor model, so the comparison targets information visibility and routing rather than a larger model or an unrelated retrieval stack.

\subsection{Research gap}
Across these threads, two gaps recur. Tutoring and educational-technology papers tend to report final-answer preference or, more rarely, human learning outcomes, while under-specifying the decision trace that produced each response. Conversely, agent and LLM papers emphasize rich traces but rarely include a learner-state ablation that isolates the value of observing the learner's own explanation. ExplainRoute connects the two perspectives as a pre-deployment audit: it logs role-level decisions under a machine-checkable contract, tests five deliberately isolating conditions, triangulates rubric scores with blinded teaching assistants, and reports when adaptive routing fails to beat simpler strategies. The experiment is intentionally narrower than a classroom trial, which lets us audit information boundaries and failure modes before claiming educational effectiveness.

\section{Methodology}
\subsection{Task and information boundary}
\begin{figure*}[t]
\centering
\includegraphics[width=0.56\textwidth]{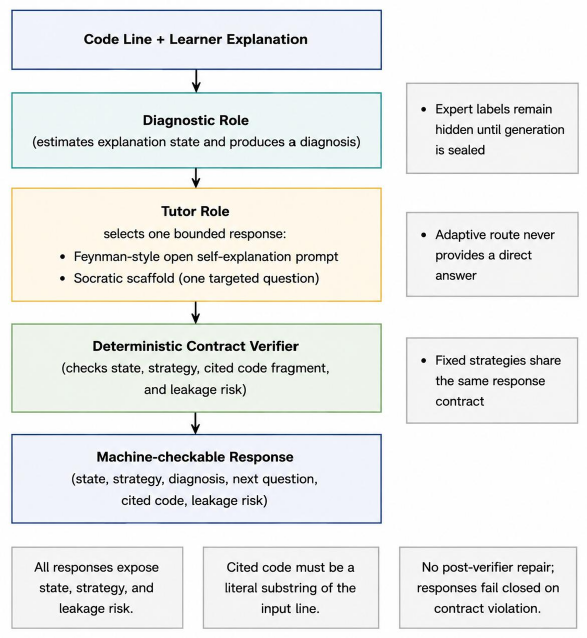}
\caption{Auditable protocol and contract boundary. The adaptive route uses a diagnostic role followed by a bounded tutor role; the verifier checks structural fields---state, strategy, answer polarity, and leakage vocabulary---before the response is accepted. Cited-code locality is audited as a separate post-generation endpoint, not as a contract condition. Expert labels remain hidden until generation is sealed.}
\label{fig:workflow}
\end{figure*}

Each example consists of one Java code line, a learner explanation, an expert explanation, and an expert annotation score from 1 to 5. Generation receives only the code line and the learner explanation. We treat this information boundary itself as part of the design under audit: restricting generation to these two fields is what makes the no-state ablation meaningful, because hiding the learner explanation from the tutor then becomes a controlled change rather than an unspecified difference between conditions. Expert explanations and scores are loaded only after the generation outputs and endpoint logs have been hashed into a generation seal, so that no expert information can influence either a generated response or its hash. No LLM-generated student answer is used as a label at any stage. Fig.~\ref{fig:workflow} overviews the protocol and the contract boundary that enforces it.

\subsection{Agent roles and response strategies}
\begin{figure*}[t]
\centering
\includegraphics[width=0.62\textwidth]{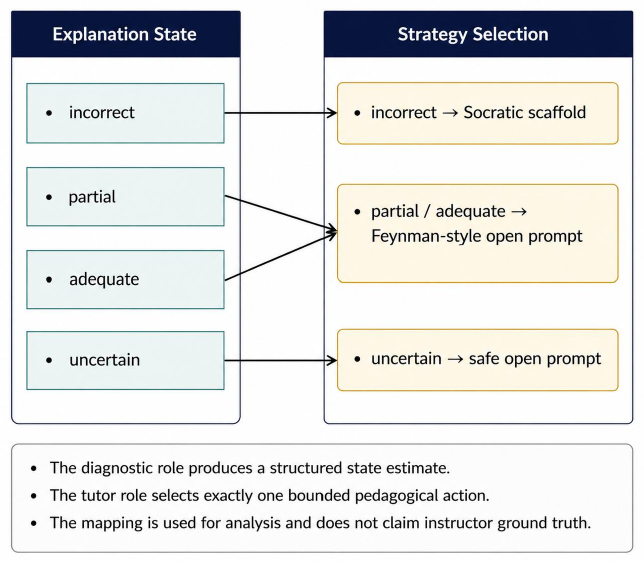}
\caption{Score-derived reference routing rule used for analysis. The diagnostic role produces a structured state estimate; the map from proxy state to a reference strategy (incorrect $\to$ Socratic; partial/adequate $\to$ open; uncertain $\to$ safe open) is a frozen analysis reference, not a hard constraint on the tutor. In the adaptive condition the tutor selects one bounded action freely, informed by---but not bound to---the diagnostic recommendation.}
\label{fig:state_strategy}
\end{figure*}

The diagnostic role, implemented with \texttt{xhang\_code}, returns one of \{\texttt{incorrect}, \texttt{partial}, \texttt{adequate}, \texttt{uncertain}\}, a short evidence-grounded diagnosis, and a recommended strategy. The tutor role, implemented with \texttt{xhang\_llm\_qwen3-235b-a22b}, receives the code, learner explanation, and diagnostic record. It selects exactly one strategy:
\begin{itemize}
\item \textbf{Feynman-style open self-explanation}: ask the learner to reconstruct code behavior or explain a small change in their own words;
\item \textbf{Socratic scaffold}: ask one bounded question targeting the likely missing relation without stating the answer.
\end{itemize}
The adaptive route cannot provide an answer. A direct-answer condition is an answer-giving baseline. The no-state ablation hides the learner explanation, marks the state uncertain, and uses a safe open prompt. Fig.~\ref{fig:state_strategy} summarizes the score-derived reference rule used for the frozen proxy analysis. In the adaptive condition the tutor freely selects one of the two bounded strategies, informed by the diagnostic record but not bound to it; the reference rule is used only to measure agreement, not to force the tutor's choice.

\subsection{Formal interface and stopping rule}
Let $c$ denote a code line and $e$ a learner explanation. The diagnostic role computes
\begin{equation}
d=f_{\mathrm{code}}(c,e), \qquad d=(s,\hat r,z),
\end{equation}
where $s$ is a four-valued state, $\hat r$ is a recommended strategy, and $z$ is a diagnosis. The tutor computes $y=f_{\mathrm{tutor}}(c,e,d)$, where $y$ is a structured response. A deterministic verifier $V(y,k)$ checks condition-specific contract $k$ and either accepts $y$ or emits a fail-closed record. There is no repair loop after verification. Human decisions are limited to the pre-run dataset, endpoint, prompt, and analysis specification; no response is edited.

\subsection{Contract and trace schema}
Every response carries the fields \texttt{state}, \texttt{strategy}, \texttt{diagnosis}, \texttt{next\_question}, \texttt{answer}, \texttt{cited\_code}, and \texttt{leakage\_risk}. The schema is shared across all five conditions so that the verifier applies one rule set rather than five ad-hoc checks: scaffold conditions require a nonempty question and an empty answer, the direct condition requires a nonempty answer, and the cited fragment must be a literal substring of the input line. Sharing the schema is precisely what allows Direct, Open, Socratic, Adaptive, and No-State to be compared on identical, machine-checkable terms. Traces retain input hashes, role decisions, token usage, latency, endpoint status, and raw-response hashes, which together support both row-level audit and aggregate cost accounting.

\begin{table*}[t]
\centering
\caption{Trace fields and audit purpose.}
\label{tab:trace}
\small
\begin{tabular}{p{0.20\textwidth}p{0.28\textwidth}p{0.39\textwidth}}
\toprule
Field family & Example fields & Purpose\\
\midrule
Input binding & explanation-record ID, code-line group ID, SHA-256 code/explanation hashes & Binds each response to an exact input without redistributing raw learner text.\\
Role decision & state, recommended strategy, diagnostic latency & Exposes adaptive decisions and separates diagnostic cost from tutor cost.\\
Tutor contract & diagnosis, question, answer, cited code, leakage flag & Enables condition-specific polarity checks; cited-code locality is a separate post-generation endpoint.\\
Operations & model, request/response hashes, usage, latency, HTTP status & Supports cost accounting and failure attribution.\\
Human gate & protocol version, no-edit marker, generation status & Distinguishes pre-run design decisions from per-example intervention.\\
\bottomrule
\end{tabular}
\end{table*}

\subsection{Prompt construction and protocol controls}
All model calls use temperature zero. The main tutor budget is 384 output tokens and the diagnostic budget is 192. The independent judge uses 768 output tokens because its served model emits a hidden reasoning preamble before the visible rubric JSON. The screened \texttt{BAAI/bge-m3} and \texttt{BAAI/bge-reranker-v2-m3} endpoints are not part of the primary run: retrieval would confound state routing with evidence acquisition.

The prompts are built from a fixed contract template and a condition-specific instruction. The system message first defines the role and output schema, then states the non-answer-giving or answer-giving rule. The user message contains the literal code line and, except in No-State, the learner explanation. The adaptive tutor additionally receives a serialized diagnostic record; the tutor is not required to copy the diagnostic recommendation, which allows us to measure routing agreement rather than merely enforce it. The diagnostic role is told to compare the explanation with the literal line and not to provide the answer.

The verifier runs after JSON extraction and enforces structural response polarity: allowed state and strategy values, a nonempty diagnosis, condition-specific answer polarity (empty answer for scaffolds and a nonempty answer for Direct), and a valid leakage-risk label. It does not inspect cited-code locality or semantic leakage, which are measured as separate post-generation endpoints. A malformed JSON object, a missing answer in Direct, or a nonempty answer in a scaffold condition produces a fail-closed record. This makes the contract independent of the judge and prevents a judge from deciding whether a response is structurally valid; it enforces structural polarity, not semantic leakage safety, which the rubric judge evaluates separately.

\begin{table}[t]
\centering
\caption{Prompt and verifier controls.}
\label{tab:prompt}
\scriptsize
\begin{tabular}{p{0.31\columnwidth}p{0.57\columnwidth}}
\toprule
Control & Fixed behavior\\
\midrule
Visible input & Code line and learner explanation, except No-State.\\
Hidden input & Expert explanation and score are never sent during generation.\\
Strategy rule & Open asks for reconstruction; Socratic asks one targeted question.\\
Answer rule & Scaffold answer field is empty; Direct answer field is nonempty.\\
Evidence rule & Cited fragment must occur literally in the code line.\\
Stopping rule & One diagnostic call at most; one tutor call; no post-verifier repair.\\
\bottomrule
\end{tabular}
\end{table}

\subsection{Execution isolation}
The holdout generator writes generation outputs and endpoint-call hashes to a new run directory. It does not load the expert-annotation file. After all five conditions for all 443 holdout explanation records finish, it writes a run manifest and seal containing artifact hashes. Only the deterministic analyzer reads the holdout annotations. The independent judge is run in a separate directory after sealing and sees the generated parsed response, code line, learner explanation, and condition. Its prompt does not contain the expert explanation, annotation score, or the fixed reference rule used for state analysis. This ordering is important because the judge is an LLM and could otherwise leak the gold state back into the generation process.

The run is parallelized across eight workers for throughput, but outputs are restored to explanation-record/condition order before hashing. The endpoint log uses a lock to preserve one JSON record per attempt. The generated artifacts therefore support both operational inspection and deterministic alignment between each explanation record and its five condition-level responses. No repeated run with a changed tutor prompt is used to replace an unfavorable result.

\subsection{Allowed actions and invariants}
The agent is not an unconstrained conversational system in this experiment. Its action space is deliberately small. The diagnostic role may classify the explanation and recommend a strategy, but it may not answer the programming question. The tutor may quote a code fragment, describe the observed explanation state, and ask one question; in the two pedagogical conditions it may not provide a worked interpretation. The verifier may reject a response, but it may not rewrite it or choose a new strategy. These restrictions make the route auditable and make the fixed baselines comparable.

Three invariants are checked independently of model confidence. First, \emph{input locality}: the cited fragment must be copied from the supplied line. Second, \emph{response polarity}: the answer field is empty for Open, Socratic, Adaptive, and No-State, and nonempty for Direct. Third, \emph{single-step interaction}: the response contains at most one student-facing question; a long list of hints is treated as a contract violation in the prompt and parser. The invariants do not prove factual correctness, but they prevent the most direct forms of evidence drift and answer leakage.

\begin{table}[t]
\centering
\caption{Auditable invariants and measurable tests.}
\label{tab:invariants}
\scriptsize
\setlength{\tabcolsep}{3pt}
\begin{tabular}{@{}p{0.27\columnwidth}p{0.48\columnwidth}p{0.16\columnwidth}@{}}
\toprule
Invariant & Test & Evidence\\
\midrule
Input locality & \texttt{cited\_code} is a substring of input code & cite rate\\
Response polarity & Answer field matches condition & contract\\
Single-step prompt & Non-direct response has one question field & question\\
Annotation separation & Seal precedes expert-annotation access & seal hash\\
No manual repair & Failed status and no-edit marker are retained & failure count\\
\bottomrule
\end{tabular}
\end{table}

\subsection{Human intervention boundary}
Human researchers define the experiment but do not act as hidden tutors. Before generation they select the public dataset, group split, models, prompt templates, output schema, token budgets, judge sample, and bootstrap seeds. During generation they do not inspect or edit individual outputs. After sealing, the analyzer reads the labels and computes fixed metrics. The only prospective intervention was an evaluator-interface repair: the first judge run was preserved because its 256-token budget truncated every JSON object, and the same predeclared rows were rerun with a larger budget. This distinction matters because an evaluator repair must not become an unreported tutor prompt search.

\subsection{Design alternatives}
We considered adding BGE-M3 retrieval, a separate reranker, and a critic-revision loop. They were excluded from the primary experiment for a causal reason: the target question is whether learner explanations provide enough information for strategy routing. Retrieval would change the available code evidence, while a critic would change response length and cost. These components can be evaluated in a future factorial experiment, but including them here would make it impossible to attribute a difference to Feynman/Socratic routing. The current design therefore prioritizes interpretability over maximal system complexity.

\section{Experiment}
\subsection{Dataset and group split}
SelfCode contains 1,770 learner--expert explanation pairs collected from ten Java programming examples \cite{chapagain2023selfcode}. Each pair is a single learner explanation of one Java code line, its corresponding expert explanation, and an expert semantic-similarity score from 1 to 5. We exclude incomplete records. After this filtering, the corpus contains 57 distinct code lines; multiple learner explanations can refer to the same line, so we use the code line as the grouping unit. The frozen split uses seed \texttt{20260725-selfcode-v1}: 46 code-line groups (1,327 explanation pairs) form development data and 11 code-line groups (443 pairs) form the untouched holdout. All explanation pairs for a given code line remain in the same partition, preventing near-duplicate code contexts from crossing the split. The holdout is skewed toward lower similarity scores: scores 1, 2, 3, 4, and 5 occur 111, 154, 108, 63, and 7 explanation pairs, respectively.

\begin{table}[t]
\centering
\caption{Distribution of expert similarity scores in the untouched holdout.}
\label{tab:dataset}
\small
\begin{tabular}{lrrrrrr}
\toprule
Similarity\\score & 1 & 2 & 3 & 4 & 5 & Total\\
\midrule
Explanation pairs & 111 & 154 & 108 & 63 & 7 & 443\\
Percent & 25.1 & 34.8 & 24.4 & 14.2 & 1.6 & 100.0\\
\bottomrule
\end{tabular}
\end{table}

The expert similarity score measures semantic agreement with the expert explanation; it is not an instructor-assigned misconception or pedagogical-strategy label. For one analysis-only proxy, we map scores 1--2 to \texttt{incorrect}, 3 to \texttt{partial}, and 4--5 to \texttt{adequate}. We then define a fixed reference rule---not ground truth for the best teaching action---that maps \texttt{incorrect} to Socratic scaffolding and \texttt{partial}/\texttt{adequate} to open self-explanation. This score-to-state mapping is an analysis assumption, not a newly observed learner label.

\subsection{Conditions and baselines}
\begin{figure*}[t]
\centering
\includegraphics[width=0.66\textwidth]{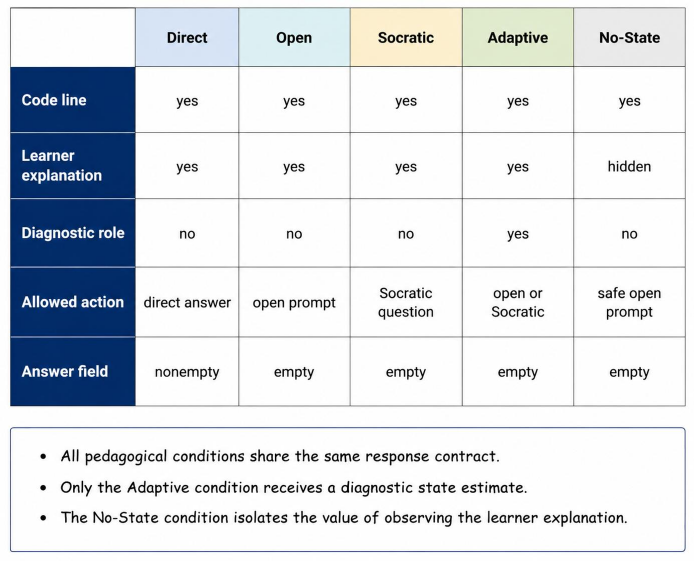}
\caption{Condition matrix and response contract. Direct, Open, Socratic, Adaptive, and No-State share the same machine-checkable response schema, but differ in input visibility, diagnostic access, allowed action, and answer-field requirements.}
\label{fig:conditions}
\end{figure*}

Fig.~\ref{fig:conditions} summarizes the five experimental conditions and the information available during generation. The baselines are chosen to answer separate audit questions rather than to create a large model leaderboard. Direct is an answer-giving baseline: it measures what is gained in apparent response quality when learner reconstruction is replaced by an explanation. Fixed Open and Fixed Socratic hold the pedagogical action constant, respectively testing open self-explanation and one-question scaffolding without routing uncertainty. No-State hides the learner explanation but retains the code line, providing the information-visibility ablation. Adaptive receives the learner explanation and one diagnostic record, then selects one of the two bounded actions; it tests the value of routing itself. All tutor conditions use the same main model and shared machine-checkable schema; Adaptive adds one \texttt{xhang\_code} call. The independent judge is \texttt{MiniMax/MiniMax-M2.7}; it receives no expert explanation or annotation score. Thus, the five conditions separate answer giving, fixed pedagogy, learner-information visibility, and adaptive routing.

\subsection{Serving configuration and reproducibility}
The model services were deployed and operated on Beihang University internal infrastructure and accessed through an OpenAI-compatible API gateway. The reported names are deployment aliases: the main tutor was \texttt{xhang\_llm\_qwen3-235b-a22b}, the code diagnostician was \texttt{xhang\_code}, and the independent judge was \texttt{MiniMax/MiniMax-M2.7}. These aliases identify the services used in the run but do not identify publicly fixed checkpoints or disclose the underlying provider/version. The main tutor and code diagnostician were screened with structured-output probes before the holdout. A probe was considered successful only when the endpoint returned HTTP 200, a nonempty message, and a parseable object under the same extraction routine used in the full run. The judge was screened separately because its hidden reasoning behavior differs from the tutor role. The run was completed on July 25, 2026; the artifact records aliases, prompts, output schemas, token budgets, request hashes, and generation seals, but not credentials or private endpoint configuration.

\begin{table}[t]
\centering
\caption{Serving and budget configuration.}
\label{tab:serving}
\scriptsize
\begin{tabular}{p{0.27\columnwidth}p{0.43\columnwidth}r}
\toprule
Role & Model alias & Max tokens\\
\midrule
Main tutor & \texttt{xhang\_llm\_qwen3-235b-a22b} & 384\\
Code diagnostician & \texttt{xhang\_code} & 192\\
Independent judge & \texttt{MiniMax/MiniMax-M2.7} & 768\\
\bottomrule
\end{tabular}
\end{table}

The parallel runner uses eight workers only to reduce wall-clock time; it does not change the number of calls per condition. Each request has a bounded timeout and up to two retries after the first attempt. Retry metadata is logged separately from successful calls. If retries still fail, the response is fail-closed rather than replaced by a fallback model. This policy is intentionally conservative because a fallback model would change the condition and make the failure rate incomparable.

\subsection{Metrics and statistical analysis}
Explanation records that refer to the same code line can share syntax, vocabulary, or response templates. Treating them as independent in a confidence interval would overstate the effective sample size. We therefore use the code-line group as the resampling unit while retaining all records within each sampled group. Point estimates and bootstrap replicates use the same record-weighted condition mean; the reported interval is the 2.5th--97.5th percentile of 100,000 code-line-cluster resamples. This avoids silently giving a code line with few records the same weight as a code line with many records.

The state endpoint is imbalanced, so we report both state accuracy and macro-F1. Accuracy describes the proportion of explanation records whose predicted state matches the similarity-score-based proxy state; macro-F1 gives each of the three proxy states equal weight. The strategy endpoint is evaluated only for Adaptive because the fixed baselines are defined to use a particular strategy. Contract validity and exact citation are binary engineering endpoints; they are not conflated with the independent judge's natural-language quality scores. Finally, the composite judge score is reported only as a summary after the five dimensions are shown, preventing a high grounding score from masking poor actionability or leakage safety.
Primary deterministic endpoints are state macro-F1, adaptive strategy accuracy, contract validity, and exact code citation. Secondary endpoints include state accuracy, question/answer presence, leakage labels, latency, token use, and endpoint failures. The independent judge scores grounding, diagnosis, actionability, leakage safety, and strategy adherence from 1 to 5; the composite is their arithmetic mean.

For a condition $m$, macro-F1 averages the three proxy-state classes. An \texttt{uncertain} prediction is an error but is not added as an absent fourth proxy class. Contract validity is the proportion accepted by $V$. All 443 holdout explanation records are reported. Paired comparisons use 100,000 code-line-group bootstrap resamples; binary contract outcomes additionally use Wilson intervals. A predeclared SHA-256 rule selects 93 explanation records (20.99\%) for blind judge evaluation, producing 465 condition-level evaluations across the five conditions. No success threshold or subset is selected after observing outputs.

\subsection{Results: state recognition and contract safety}
\begin{figure*}[t]
\centering
\includegraphics[width=0.85\textwidth]{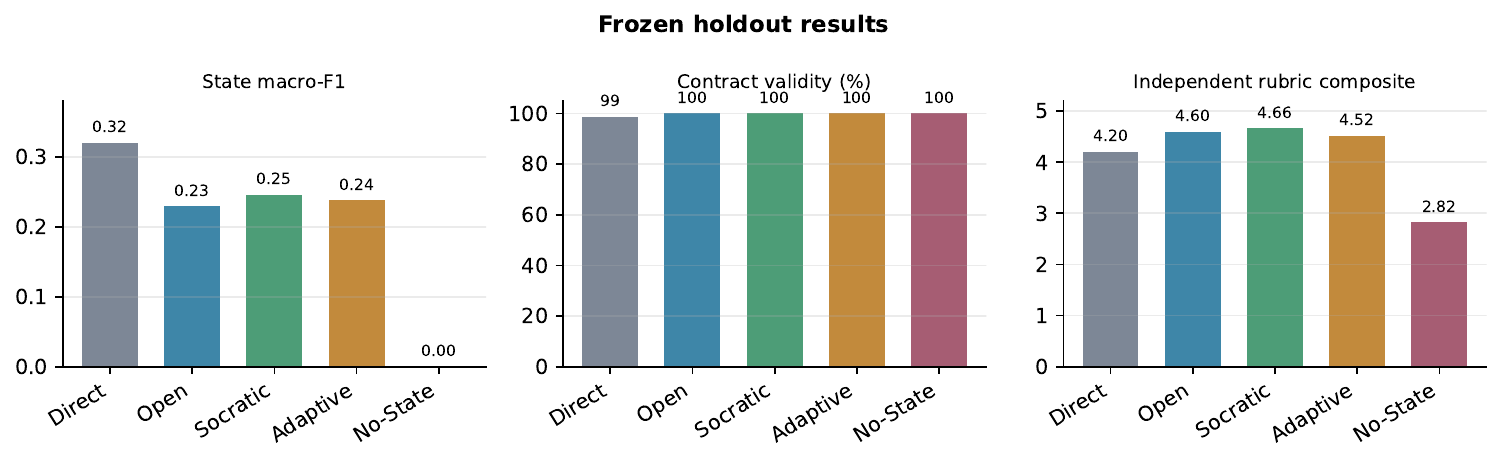}
\caption{Frozen holdout overview. Contract validity is formal; state macro-F1 is agreement with the proxy state derived from expert similarity scores; the independent composite is an observable-response rubric. Adaptive is safer and more capable than No-State, but fixed strategies remain competitive.}
\label{fig:overview}
\end{figure*}

Fig.~\ref{fig:overview} provides the visual summary, and Table~\ref{tab:results} consolidates the deterministic and judge endpoints. Fixed Open and Socratic responses satisfy the scaffold contract on all rows. Adaptive routing also satisfies the contract on all 443 explanation records and reaches 60.5\% agreement with the fixed reference-rule route. Its state macro-F1 is 0.238, so the adaptive diagnosis remains weak even though the final response is structurally safe. No-State is intentionally unable to match any similarity-score-based proxy state. Adaptive and No-State differ in more than learner-explanation visibility (Adaptive also adds a diagnostic call and strategy freedom), so the cleaner visibility isolation is Fixed Open versus No-State---same tutor, open strategy, no diagnostic---which we report with the paired comparisons below (state accuracy $+0.278$; judge composite $+1.778$).

\begin{table*}[t]
\centering
\caption{Consolidated holdout results. Deterministic contract, state, route, and citation values use all 443 holdout explanation records; judge dimensions use the predeclared 93-record blind sample. Contract validity, state accuracy, reference-rule route agreement, and exact citation are percentages.}
\label{tab:results}
\scriptsize
\setlength{\tabcolsep}{3.4pt}
\begin{tabular}{@{}lccccc@{\quad}rrrrrr@{}}
\toprule
& \multicolumn{5}{c}{Deterministic holdout} & \multicolumn{6}{c}{Independent judge}\\
\cmidrule(lr){2-6}\cmidrule(l){7-12}
Condition & Contract & State & F1 & Route & Cite & Ground & Diagn. & Action & Safety & Adhere & Comp.\\
\midrule
Direct & 98.65 & 32.05 & .320 & -- & 98.65 & 4.613 & 4.161 & 2.828 & 4.828 & 4.591 & 4.204\\
Open & 100.00 & 27.77 & .229 & -- & 99.55 & 4.355 & 4.452 & 4.505 & 4.978 & 4.699 & 4.598\\
Socratic & 100.00 & 27.09 & .246 & -- & 99.32 & 4.452 & 4.484 & 4.624 & 4.935 & 4.796 & 4.658\\
Adaptive & 100.00 & 23.02 & .238 & 60.50 & 100.00 & 4.398 & 4.398 & 4.462 & 4.957 & 4.366 & 4.516\\
No-State & 100.00 & 0.00 & .000 & -- & 100.00 & 2.215 & 1.989 & 2.634 & 4.688 & 2.570 & 2.819\\
\bottomrule
\end{tabular}
\end{table*}

\subsection{Results: independent response rubric}
The blind judge rates Adaptive above No-State by 1.697 composite points (95\% CI [1.569,1.956]), showing the value of observing the learner explanation. Relative to Open, the Adaptive composite difference is $-0.082$ (CI [$-0.217,-0.019$]); relative to Socratic, it is $-0.142$ (CI [$-0.204,0.036$]). The five judge dimensions in Table~\ref{tab:results} show that No-State is weakest in grounding, diagnosis, actionability, and adherence. The direct baseline remains grounded but has low actionability because it gives an answer instead of eliciting reconstruction.

\subsection{Routing behavior, cost, and failure analysis}
Adaptive selected Socratic for 336/443 explanation records (75.8\%) and Open for 107/443 (24.2\%). Every Adaptive output declared low leakage risk, and all 443 code fragments were exact substrings. Operationally, Adaptive costs one additional diagnostic call: its mean latency is 6.14 seconds, compared with 4.31 seconds for Direct, 3.89 seconds for Open, 3.63 seconds for Socratic, and 3.35 seconds for No-State. Adaptive uses two calls per row; every other condition uses one. Mean question lengths are 20.17 words for Open, 14.80 for Socratic, 17.29 for Adaptive, and 15.66 for No-State, while Direct answers average 36.79 words. Within Adaptive, the diagnostic role has a median latency of 1.93 seconds and the tutor role a median latency of 3.79 seconds, making the additional call a measurable cost of state-aware routing. The independent judge calls are excluded from tutor latency and reported in a separate manifest. Token usage is retained in endpoint logs, but aggregate claims in this paper focus on calls and latency because gateway token accounting can vary across model providers.

\begin{figure}[t]
\centering
\includegraphics[width=\columnwidth]{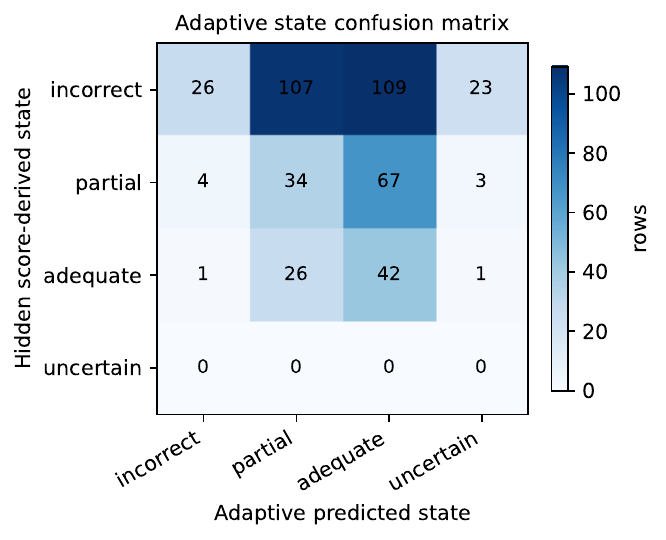}
\caption{Adaptive state confusion matrix. Low-score explanations are frequently mapped to partial or adequate, motivating calibration and richer supervision.}
\label{fig:confusion}
\end{figure}

The generation endpoint had no HTTP failures. Six of 2,215 tutor traces (0.27\%) failed closed because the direct-answer response omitted the required answer field; they remain in the denominator. The first judge run had 465/465 parse failures because a 256-token budget stopped before JSON. We preserve that interface failure, increase only the judge budget to 768 tokens, and obtain 465/465 parsed outputs in the corrected run. No response was manually edited.

\subsection{Paired comparisons and interpretation}
The record-weighted Adaptive--Open state-accuracy difference is $-0.047$ (95\% code-line-cluster bootstrap CI [$-0.100,0.006$]); Adaptive--Socratic is $-0.041$ (CI [$-0.096,0.008$]); Adaptive--No-State is $+0.230$ (CI [$0.169,0.321$]). The corresponding judge composite differences are $-0.082$ (CI [$-0.186,-0.007$]), $-0.142$ (CI [$-0.241,-0.032$]), and $+1.697$ (CI [$1.459,1.976$]). Because Adaptive and No-State confound learner-explanation visibility with a diagnostic call and strategy freedom, the cleaner visibility isolation is Fixed Open versus No-State (same tutor, open strategy, no diagnostic): state accuracy $+0.278$ (CI [$+0.212,+0.362$]) and judge composite $+1.778$ (CI [$+1.508,+2.106$]) in favor of showing the learner explanation. The information-value claim rests on this cleaner comparison, not on the confounded Adaptive--No-State pipeline difference. These are offline proxy comparisons, not causal effects.

\begin{table*}[t]
\centering
\caption{Paired record-weighted Adaptive differences. Intervals are 95\% code-line-cluster bootstrap intervals; each sampled code-line group contributes all of its explanation records.}
\label{tab:paired}
\small
\begin{tabular}{lrrl}
\toprule
Comparison & $\Delta$ state accuracy & $\Delta$ judge composite & Interpretation\\
\midrule
Adaptive $-$ Open & $-0.047$ [$-0.100,0.006$] & $-0.082$ [$-0.186,-0.007$] & Open remains competitive\\
Adaptive $-$ Socratic & $-0.041$ [$-0.096,0.008$] & $-0.142$ [$-0.241,-0.032$] & No adaptive advantage\\
Adaptive $-$ No-State & $+0.230$ [$0.169,0.321$] & $+1.697$ [$1.459,1.976$] & Pipeline above no-state\\
Open $-$ No-State & $+0.278$ [$+0.212,+0.362$] & $+1.778$ [$+1.508,+2.106$] & Visibility isolates value\\
\bottomrule
\end{tabular}
\end{table*}

\subsection{State confusion and route calibration}
The confusion matrix in Fig.~\ref{fig:confusion} reveals a systematic calibration problem rather than random parsing noise. Among 265 examples assigned the similarity-score-based proxy state \texttt{incorrect}, the adaptive role predicts \texttt{partial} 107 times, \texttt{adequate} 109 times, and \texttt{uncertain} 23 times; only 26 are predicted as \texttt{incorrect}. For the 108 proxy-\texttt{partial} examples, 67 are mapped to \texttt{adequate}; for the 70 proxy-\texttt{adequate} examples, 42 are correctly mapped. Thus, the diagnostic role tends to soften low-coverage explanations instead of preserving the distinction needed by the fixed reference rule. This is why 60.5\% reference-rule route agreement can coexist with a 0.238 state macro-F1.

The fixed strategies provide a useful calibration reference. Open predicts \texttt{partial} on most examples because its instruction is a response mode rather than a state classifier; Socratic behaves similarly. Their state metrics are not evidence that they understand the learner, but they show that the adaptive state estimate does not automatically translate into a better final question. Future versions should calibrate state probabilities or train on instructor-labeled misconceptions instead of relying on a four-valued free-form diagnosis.

\subsection{Independent judge protocol and evaluator risk}
The blind judge evaluates five observable properties separately: grounding, diagnosis, actionability, leakage safety, and strategy adherence. It receives the condition name so that adherence is testable, but it cannot compare the response with the hidden expert explanation. The 93-record sample is selected by a hash of the explanation-record identifier before expert annotations are loaded; all five conditions for a selected record are judged, which creates paired comparisons and avoids a favorable-condition sample.

The judge is intentionally treated as an evaluator proxy rather than an authority. The first judge run demonstrates why this matters: a 256-token budget caused every output to terminate before the requested JSON. We retain the failed run and rerun the same 465 condition-level evaluations with 768 tokens, obtaining 465 parseable outputs. This is an interface-reliability repair, not a second chance to improve the tutor response. The paper reports both the failure and the corrected result.

\subsection{Blinded teaching-assistant evaluation}
To test whether the independent LLM judge is a valid evaluator, three teaching assistants for an introductory Java course rated a stratified subset of 40 explanation records from the frozen blind sample under the same five-dimension rubric (200 condition-level responses: 40 records $\times$ 5 conditions). Each rater saw exactly what the judge saw---code line, learner explanation, condition label, and generated response---and none of the expert fields, and rated independently after a non-holdout calibration session. The selected explanation records, rubric, and analysis were frozen before rating (\texttt{experiment/HUMAN\_EVAL\_PROTOCOL.md}, \texttt{RUBRIC\_GUIDE.md}).

Inter-rater reliability is bounded by ceiling effects. The composite reached a mean pairwise quadratic-weighted $\kappa = 0.32$ (fair). On the two dimensions where responses varied, agreement was usable (actionability $\kappa = 0.71$, diagnosis $\kappa = 0.57$); on near-saturated dimensions (grounding, leakage safety, strategy adherence, all with means 4.8--4.9 out of 5), absolute agreement was high (78--87\% exact; $\geq$99\% within one point) but $\kappa$ was uninformative because of restricted variance. We therefore treat the human composite as a corroborating signal rather than a precision instrument.

On the 200 shared items the TA and LLM-judge composites correlated at Spearman $\rho = 0.57$ (mean absolute disagreement 0.55 points); humans scored $+0.27$ higher on average, largely because they were more lenient on the No-State ablation (human composite 3.92 vs.\ judge 2.82). The three headline comparisons, recomputed with the TA composite under the same 11-group bootstrap, corroborate the central result. The information-value finding replicates: Adaptive exceeds No-State by $+0.74$ composite points (95\% CI $[+0.71,+0.77]$), smaller than the judge's $+1.70$ but the same sign with a CI far above zero. Under human rating Adaptive is statistically tied with the fixed strategies---$+0.05$ vs.\ Open (CI $[-0.01,+0.11]$) and $+0.01$ vs.\ Socratic (CI $[-0.07,+0.07]$)---where the LLM judge had placed it slightly below; both evaluators therefore agree there is no reliable adaptive gain over the fixed strategies. Adaptive also leads Direct by $+0.43$ (CI $[+0.38,+0.48]$).

These results corroborate the LLM judge on the central information-value claim and on the absence of an adaptive advantage over fixed strategies, while bounding the strength of either evaluator: human and LLM agree on ranking but differ in leniency, and inter-rater reliability is only fair. The study remains an offline rubric comparison; it does not measure student learning.

\begin{table*}[t]
\centering
\caption{Representative audited Adaptive cases from the blinded TA subset. TA values are means across the three raters on the five rubric dimensions; they are illustrative cases, not additional aggregate endpoints.}
\label{tab:cases}
\scriptsize
\setlength{\tabcolsep}{3pt}
\begin{tabular}{@{}p{0.11\textwidth}p{0.29\textwidth}p{0.30\textwidth}p{0.22\textwidth}@{}}
\toprule
Case & Learner explanation and code line & Audit trace & Human rubric\\
\midrule
\auditcase{1006} & ``This is the continuation test. do-while condition & proxy state: incorrect; predicted state: partial; route: Socratic. Question: what happens when \texttt{num>0} is false? & Ground 4.7; Diagn. 2.7; Action 4.7; Safety 5.0; Adhere 5.0\\
\auditcase{1229} & ``Initialize a max value ... by taking the first value. \texttt{int maxValue=values[0];} & proxy state: incorrect; predicted state: adequate; route: Socratic. Question: why initialize with the first element? & Ground 5.0; Diagn. 4.0; Action 5.0; Safety 5.0; Adhere 4.7\\
\bottomrule
\end{tabular}
\end{table*}

The cases illustrate why the framework reports both safety and state calibration. In
Case~\auditcase{1006}, the route remains non-answer-giving and highly actionable while
human raters judge the diagnosis more cautiously. In Case~\auditcase{1229}, the response
is safe and useful as a Socratic prompt, but the state estimate is too optimistic relative
to the similarity-score-based proxy (the record is mapped from proxy-\texttt{incorrect} to predicted \texttt{adequate}). The audit therefore separates
a safe pedagogical action from a reliable learner-state estimate; a response can satisfy
the former while exposing a failure in the latter.

\subsection{Validity, ethics, and threats to validity}
SelfCode is a small Java-only sentence-level corpus; its score-to-state mapping is a proxy, not a complete learner model. The independent judge is another language model rather than a human instructor. The experiment has no student intervention, post-test, retention measure, or course outcome. It cannot establish instructional effectiveness. Public availability does not prove that model pretraining was uncontaminated. Raw student text, private endpoints, credentials, and unredacted traces are excluded from the release.

The authors fixed the corpus version, code-line grouping rule, split seed, conditions, model aliases, prompts, budgets, judge sampling rule, and bootstrap seeds before reading the holdout expert annotations. The agent decided state, strategy, diagnosis, question, cited code, and leakage flag; a deterministic parser decided contract validity. The generation seal reports that hidden labels were not read before generation and its artifact hashes match 2,215 condition traces and 2,658 model-role calls. The release bundle contains the protocol, scripts, aggregate row metrics, judge analysis, and figures, while the raw corpus is referenced rather than redistributed.

\textbf{Construct validity.} SelfCode scores semantic similarity to an expert explanation, whereas a tutor may need to recognize uncertainty, goal-level reasoning, or misconceptions that are not paraphrases. The three-state mapping also introduces thresholds. The fixed reference rule is thus a diagnostic reference, not ground truth for the best pedagogical action. Similarly, the independent judge scores observable properties and does not replace a trained instructor rubric.

\textbf{Internal validity.} The same main tutor model is used across conditions, but Adaptive receives an additional diagnostic message. The no-state ablation hides the learner explanation but retains the code line, so its difference is attributable to missing learner information within the frozen interface. Temperature zero reduces sampling variation but does not guarantee deterministic serving behavior. Group bootstrap treats code-line groups as exchangeable; templates shared across examples may induce residual dependence.

\textbf{External validity.} The corpus contains Java lines and sentence-level explanations from ten examples. Results may differ for Python, data structures, debugging traces, multilingual learners, long explanations, or multi-turn dialogue. The Beihang-hosted deployment aliases are not public fixed checkpoints, and the underlying provider/version metadata are not available for release. Exact weight-level replication is therefore not possible from the aliases alone. Functional replication can instead use any OpenAI-compatible deployment that provides (i) a structured-output code diagnostician, (ii) an instruction-following tutor, and (iii) an independent rubric judge, together with the released prompts, schemas, temperature-zero setting, token budgets, split, and analysis scripts; such replication may still require output-format calibration.

\textbf{Evaluator validity.} The independent judge is the same broad class of generative system as the tutor and can share stylistic biases, so we triangulate it with the blinded teaching-assistant evaluation above: the TA and LLM composites correlate at $\rho = 0.57$ and agree on the information-value result and on the absence of an adaptive advantage, though humans were more lenient and inter-rater reliability was only fair (composite $\kappa = 0.32$). Both evaluators saw the condition label (required to score strategy adherence); this could bias non-adherence dimensions through a label-halo effect, particularly for the descriptively named No-State condition, and a neutral-coding replication for the non-adherence dimensions is left to future work. The two cases in Table~\ref{tab:cases} further show why aggregate agreement is insufficient: humans can accept a response as safe and actionable while disagreeing with its state diagnosis. The five dimensions are reported separately, and the failed first judge run is retained rather than silently discarded. A future study should extend this triangulation to blinded instructors together with a learning-outcome measure.

\subsection{Deployment implications}
The contract is appropriate for a low-risk prototype in which a tutor asks a learner to explain code, but it is not a license to automate grading or educational decisions. A deployment should enforce data minimization, serving-layer retention controls, redaction, and a human escalation path. In particular, a question that avoids the explicit answer can still reveal the answer through its wording. The current system should therefore be used as an auditable response generator whose traces are reviewable, not as an autonomous teacher whose output is accepted without oversight.

\section{Conclusion}
We presented \system, a two-role teaching-assistant agent that keeps a programming tutor's decisions inspectable: the diagnostic role emits a structured explanation-state estimate, the tutor selects exactly one bounded Feynman-style or Socratic action, and a deterministic verifier checks state, strategy, cited code, and leakage risk before any response is accepted. Because every response carries the same machine-checkable contract, the system's behavior---and its failures---can be audited rather than only observed as fluent dialogue.

On a frozen, group-held-out SelfCode evaluation the results are mixed but informative. Contract validity is 100\% for the fixed pedagogical conditions and for Adaptive, confirming that a non-answer-giving tutor can be made structurally safe at scale. Observing the learner explanation has clear value: Adaptive's blind composite rubric score (4.516/5) is 1.697 points above the no-state ablation (2.819/5). At the same time, Adaptive does not exceed fixed Open (4.598) or Socratic (4.658) strategies, and its state macro-F1 of 0.238 reveals a systematic calibration problem in which low-coverage explanations are softened rather than distinguished. The honest reading is that auditable, non-answer-giving interaction is achievable, but a weak state estimate does not automatically translate into a better question. A blinded teaching-assistant rubric evaluation (three raters, 40-row subset) corroborates these conclusions: it reproduces the information-value result (Adaptive above No-State by $+0.74$ composite points under human rating) and finds Adaptive statistically tied with---rather than below---the fixed strategies, though inter-rater reliability was only fair (composite $\kappa = 0.32$).

These are offline proxy results, not evidence of learning gain, retention, or causal instructional effectiveness. To support stronger claims we see three concrete next steps: instructor-labeled misconceptions and calibrated state probabilities to address the calibration problem, multi-turn learner responses to test whether adaptive routing helps over a full dialogue rather than a single turn, and a prospective study with blinded instructor ratings and a learning-outcome measure (the present TA rubric is a step toward, but not a substitute for, this). Within its current scope, \system\ contributes an auditable interaction protocol and a measurable information-value result, and it makes explicit where adaptive routing currently falls short.

\section*{Acknowledgments}
This work was supported in part by the National Natural Science Foundation of China under Grant 62406014, in part by the Noncommunicable Chronic Diseases–National Science and Technology Major Project, 2025ZD0546500, in part by the Beijing Natural Science Foundation (7262079), in part by the Zhejiang Natural Science Foundation (LZ26F020012) and in part by the Start-up Funds of Hangzhou International Innovation Institute of Beihang University under Grant No.2024KQ045.

\bibliographystyle{IEEEtran}
\bibliography{references}

\end{document}